\documentclass[runningheads,a4paper]{article}

\usepackage{amssymb}
\usepackage{graphicx}
\usepackage{epsf}

\usepackage{url}
\urldef{\mailsa}\path|alexander.gamkrelidze@tsu.ge|
\urldef{\mailsb}\path|g.hotz@rz.uni-sb.de|    
\urldef{\mailsc}\path|varlevani@gmail.com|

\usepackage{array}
\usepackage{cite}
\usepackage{enumerate}

\begin{document}


\title{New Invariants for the Graph Isomorphism Problem}

\author{Alexander Gamkrelidze\\
alexander.gamkrelidze\\
Department of Computer Science,\\
Tbilisi State University,\\
Building XI, Room 355,\\
Tbilisi, Georgia
\and G\" unter Hotz\\
g.hotz@rz.uni-sb.de\\
University of Saarland\\
Campus E 1.1, Room 311\\
Department of Computer Science
\and Levan Varamashvili\\
varlevani@gmail.com\\
Google Inc., Z\" urich,\\
Switzerland
}

%
%

\maketitle

\begin{abstract}
In this paper we introduce a novel polynomial-time algorithm to compute graph invariants based on the modified random walk idea on graphs. However not proved to be a full graph invariant by now, our method gives the right answer for the graph instances other well-known methods could not compute (such as special F\"urer Gadgets and point-line incidence graphs of finite projective planes of higher degrees).

\

\end{abstract}

\section{Introduction}

The computational complexity of graph isomorphisms remains unsolved
for over three decades now. No polynomial-time algorithm deciding
whether two given graphs are isomorphic is known; neither could
this problem be proved to be NP-complete. Besides its practical
importance, the solution of this problem would give us a deeper
insight into the structure of complexity hierarchies (in the case
where it is NP-complete, the polynomial hierarchy would collapse
to its second level, \cite{Sch, Boppana}). The counting version of
the graph isomorphism problem is known to be reducible to its
decisional version (see~\cite{Mathon}), while for all known
NP-complete problems, their counting versions seem to be much
harder. On the other hand, the known lower bounds in terms of
hardness results for the graph isomorphism problem are
surprisingly weak. Because of such theoretical results and of many
failed attempts to develop an efficient algorithm for graph
equivalence or prove its NP-completeness, it could be one of the
intermediate problems that are neither   P nor NP-complete ---
P$\neq$ NP assumed.

Despite these facts, efficient algorithms are known for several
special classes of graphs such planar graphs \cite{Tarjan,
Hopcroft}, random graphs \cite{Erdos}, graphs with bounded
eigenvalue multiplicity \cite{Babai}, graphs of bounded genus
\cite{Filotti} and graphs of bounded degree \cite{Luks}. In some
cases, like trees \cite{Lindell, Buss}, or graphs with colored
vertices and bounded color classes \cite{Luks1}, even NC
algorithms for isomorphism exist. An interesting result is due to
Erd\"os et al.~\cite{Erdos} who showed that the isomorphism
problem is easily solvable by a naive algorithm for ``almost all''
graphs: only a small amount of special graphs make all known
algorithms impracticalble. Surprisingly, it is very hard to find
graph instances that cause problems for all known graph
isomorphism systems. Only highly regular graphs such as F\"urer
gadgets, point-line incidence graphs of finite projective planes
or graphs for Hadamard matrices cause problems to even leading
graph-isomorphism solvers such as ``Nauty'',~\cite{Nauty}.

 For each graph with $n$ vertices we build a
set of $n(n-1)/2$ quadratic polynomials that builds a graph
invariant. The upper bound of the time complexity for the
developed algorithm is $O(n^5)$, but in most cases it can be
reduced to~$O(n^4)$.

At the end, we give computational results -- tables of the running
times of our program on some difficult graphs (special F\"urer
gadgets and point-line incidence graphs of finite projective
planes of higher degrees).

\section{Basic Ideas}

We define the random walk on graphs as follows: given a graph
$G=(V,E)$ with vertex set $V$ and the set of edges $E$, if at time
$t_i$ we are in a vertex $v\in V$, at the next moment $t_{i+1}$ we
decide to move to one of the neighboring vertices or to stay in
the actual vertex (here and further in this work we consider
discrete time), the probability to stay in the actual vertex $v_i$
be $p_i$ and the probability to move from vertex $v_i$ to $v_j$ be
$p_{i,j}$. If we decide to change the vertex, the probabilities to
choose one of the neighbors are equal and we obtain the following
formula:
$$
 p_{i,j}=\left\{
 \begin{array}{ll}
 \frac{1-p_i}{d(v_i)}, &\mbox{if $(v_i,v_j)\in E$}, \\
 0 &\mbox{otherwise,}
 \end{array}
 \right.
$$
where $d(v_i)$ is the degree of $v_i\in V$.

Choosing different probabilities $p_i$ for staying in a particular
vertex $v_i$, we can vary the changing probabilities $p_{i,j}$
thus producing different directed weighted graphs that originate
from $G$ with the connection graph $M'_G=(m'_{i,j})_{i,j-1}^n$,
where $m'_{i,j}$ is the changing probability from $v_i$ to $v_j$
if $i\neq j$ and the staying probability else.

Dividing each row by $\frac{1-p_i}{d(v_i)}$ we can normalize
$M'_G$ and obtain another connection matrix
$M_G=(m_{i,j})_{i,j-1}^n$, where
$$
 m_{i,i}=X_i=\frac{p_id(v_i)}{1-p_i}
$$
and for $i\neq j$
$$
 m_{i,j}=\begin{array}{ll}
 1, & \mbox{if $(v_i,v_j)\in E$},\\
 0 & \mbox{otherwise.}
 \end{array}
$$

Thus, for a given graph $G$, we obtain a connection matrix $M_G$
as usual with variables $X_i$ on the diagonal.

Now consider a connection matrix $M^{i,j}_G(\eta)$ with fixed
values $\eta$ on the diagonal up to the positions
$m_{i,i}=m_{j,j}=X$:
$$ M_G^{2,5}(\eta)=\begin{array}{ccccccc}
 \eta & 1 & 0 & 0 & 1 & 1 & 0\\
 1 & X & 0 & 0 & 1 & 0 & 1\\
 0 & 0 & \eta & 1 & 1 & 1 & 0\\
 0 & 0 & 1 & \eta & 1 & 0 & 1\\
 1 & 1 & 1 & 1 & X & 0 & 1\\
 1 & 0 & 1 & 0 & 0 & \eta & 1\\
 0 & 1 & 0 & 1 & 1 & 1 & \eta\\
 \end{array},\quad
 A_G=\begin{array}{ccccccc}
 1 & 1 & 0 & 0 & 1 & 1 & 0\\
 1 & 1 & 0 & 0 & 1 & 0 & 1\\
 0 & 0 & 1 & 1 & 1 & 1 & 0\\
 0 & 0 & 1 & 1 & 1 & 0 & 1\\
 1 & 1 & 1 & 1 & 1 & 0 & 1\\
 1 & 0 & 1 & 0 & 0 & 1 & 1\\
 0 & 1 & 0 & 1 & 1 & 1 & 1\\
 \end{array}.
$$

Let $P_G^{i,j}(X)=det(M_G^{i,j})=|M_G^{i,j}|$ and $A_G$ be the
usual connection matrix for $G$ with the diagonal elements~1. It
is easy to verify that
$$
 M_G^{i,j}(1)=|A_G^{i,j}|X^2
 +\big(|B_G^{i,j}|+|B_G^{j,i}|\big)X+|C_G^{i,j}|,
$$
where
\begin{enumerate}[(i)]
 \item %
$A_G^{i,j}$ is the matrix $A_G$ with $i$th and $j$th raw and
column eliminated;
 \item %
$B_G^{i,j}$ is the matrix $A_G$ with $i$th raw and column
eliminated and $m_{i,i}=0$;
 \item %
$C_G^{i,j}$ is the matrix $A_G$ with $m_{i,i}=m_{j,j}=0$.
\end{enumerate}

\begin{figure}[h]
 \centering
 \includegraphics[width=0.5\textwidth]{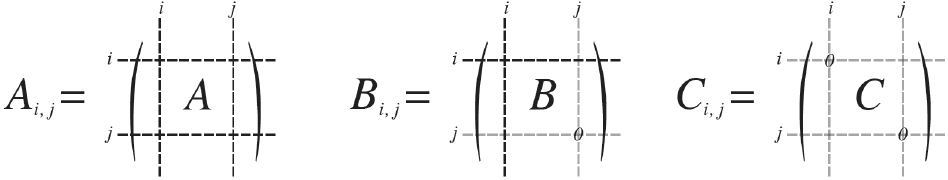}
 \caption{Figure Example.}
 \label{fig:example}
\end{figure}

Since $P_G^{i,j}(1)=|A_G|$, the polynomial $P_G^{i,j}(1)-|A_G|$
is divisible by $X-1$ and we obtain:
$$
 L_G^{i,j}(X)=\frac{P_G^{i,j}(1)-|A_G|}{X-1}
 =|A_G^{i,j}|X+\big(|A_G|-|C_G^{i,j}|\big);
$$
this can be easily verified by using the equation
$$
 |A_G|-|C_G^{i,j}|-|A_G^{i,j}|=|B_G^{i,j}|+|B_G^{j,i}|.
$$
Hence,
$$
 \mathfrak A_G=\big|\big(|A_G^{i,j}|\big)_{i,j=1}^n\big|,
 \quad
 \mathfrak C_G=\big|\big(|C_G^{i,k}|\big)_{i,j=1}^n\big|
$$
are graph isomorphisms (here and further in this work we set
 $n=|V|$ as the number of vertices of~$G$).

Note that if $\mathfrak A_G\neq\mathfrak A_G'$ or $\mathfrak
C_G\neq\mathfrak C_G'$, the graphs $G$ and $G'$ are definitely
not isomorphic. On the contrary, if $\mathfrak A_G=\mathfrak
A_G'$ and $\mathfrak C_G=\mathfrak C_G'$, these two graphs are
not proved to be equivalent. But if we regard
$\big(|A_G^{i,j}|\big)_{i,j=1}^n$ and
$\big(|C_G^{i,k}|\big)_{i,j=1}^n$ as the connection matrices for
some weighted graphs, we can repeat the whole method recursively
by setting $A_G=\big(|A_G^{i,j}|\big)_{i,j=1}^n$ and
$A_{G'}=\big(|A_{G'}^{i,j}|\big)_{i,j=1}^n$.

We summarize the above ideas in the following algorithm.

\vskip+0.2cm

\noindent {\bf Algorithm}{ \it GraphIsomorphism}

\noindent
{\bf Input:} connection matrices $A_G$ and $A_{G'}$
 of the graphs $G$ and $G'$; \\[0.2cm]
{\ttfamily
\noindent
Repeat the following code twice


\hskip.1cm $\{$

\hskip.1cm compute the matrices
 $\big(|A_G^{i,j}|\big)_{i,j=1}^n$ and
 $\big(|A_{G'}^{i,j}|\big)_{i,j=1}^n$;

If (the sets of entries of these matrices are not equal)

\hskip.4cm $G\not\cong G'$ and stop;

\hskip.2cm else If($\mathfrak A_G\neq\mathfrak A_{G'}$)

\hskip2cm $G\not\cong G'$ and stop;

\hskip1.7cm else If($\mathfrak C_G\neq\mathfrak C_{G'}$)

\hskip3cm $G\not\cong G'$ and stop;

\hskip.2cm Set
 $A_G=\big(|A_G^{i,j}|\big)_{i,j=1}^n$ and
 $A_{G'}=\big(|A_{G'}^{i,j}|\big)_{i,j=1}^n$;

\hskip.1cm $\}$

\noindent
$G\cong G'$
}
\vskip+0.2cm

Obviously, the complexity of this method is $O(n^5)$, but using
the special methods (computation of the inverse matrix of $A_G$ in
the case $\det(A_G)\neq0$ and using its symmetry) the upper bound
can be reduced to~$O(n^4)$.

Another problem that arises during the computation is that the
determinants
$$
 \big|\big(|A_G^{i,j}|\big)_{i,j=1}^n\big|,
 \quad  \big|\big(|A_G^{i,j}|\big)_{i,j=1}^n\big|
$$
can become very large. In our computations, for the projective
plane of order~16 with the $546\times546$ connection matrix each
determinant is~1086 decimal digits long (these computations were
done using Wolfram Mathematica~8). This problem can be solved by
the computations in different finite fields $\mathbb Z_{p_l}$,
$D_l=|A_G^{i,j}|\mbox{\it \ mod }{p_l}$, $l\in\{1,\dots,k\}$, for coprime
$p_i$s so that $P=p_1\cdot p_2\cdots p_k\geq |A_G^{i,j}|$.
$|A_G^{i,j}|\in\mathbb Z_P$ can be restored due to the Chinese
remainder theorem.

\section{Computations and Experimental Results}

As was shown in \cite{Erdos}, the graph isomorphism problem can be
solved efficiently for almost all graphs. The problems arise only
considering special cases. The most efficient system known for
today is Bredan McKay's Nauty \cite{Nauty}. It is very efficient
for most known hard graphs but has exponential running time on
special family of graphs of bounded degree called Miyazaki
graphs~\cite{Miyazaki}. Besides this, most known systems have
significant problems with the graphs originated from projective
planes.

In this section, we show the experimental results after applying
our methods to distinguish some types of hard graphs.

\subsection{F\"urer gadgets: Miyazaki graphs}

\begin{figure}[h]
 \centering
 \includegraphics[width=0.9\textwidth]{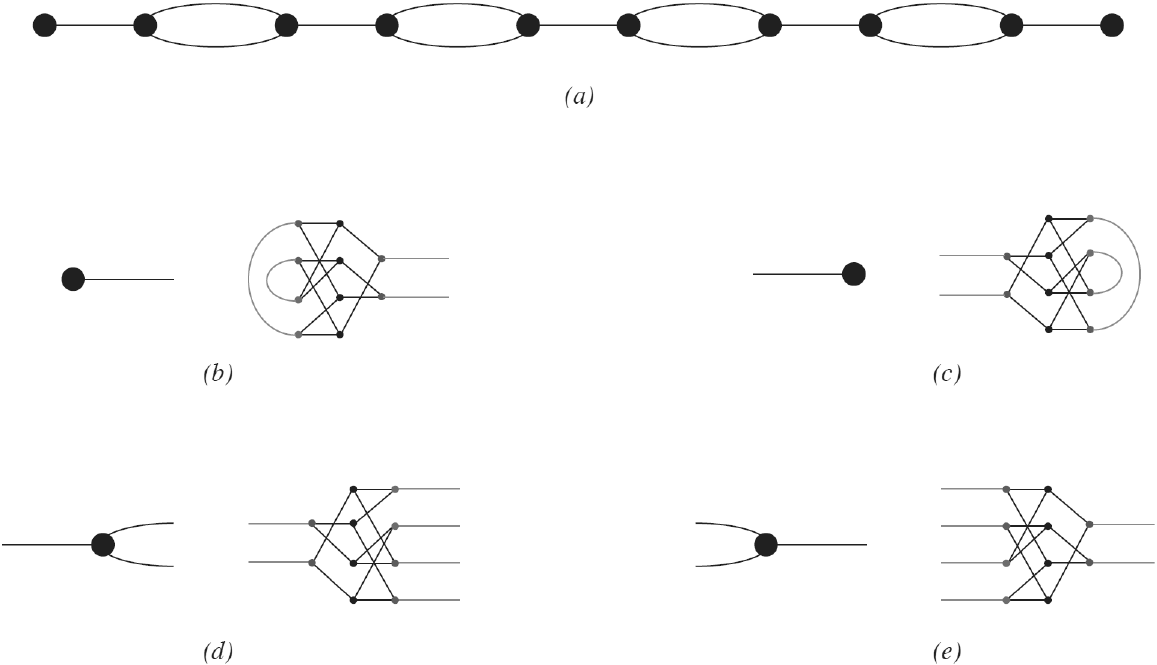}
 \caption{Example of a Miyazaki graph and its parts.}
 \label{Fig2}
\end{figure}

Miyazaki graphs are special cases of a so called
Cai--F\"urer--Immerman construction based on F\"urer gadgets
(\cite{Furer}). As an example consider a Miyazaki graph $M_4$
shown in Fig.~\ref{Fig2}(a).

The bold circles represent special subgraphs shown in
Fig.~\ref{Fig2}(b)--(e). One can easily extend this construction
to a Miyazaki graph $M_k$ with arbitrary~$k$.

\begin{figure}[h]
\centering
 \includegraphics[width=0.9\textwidth]{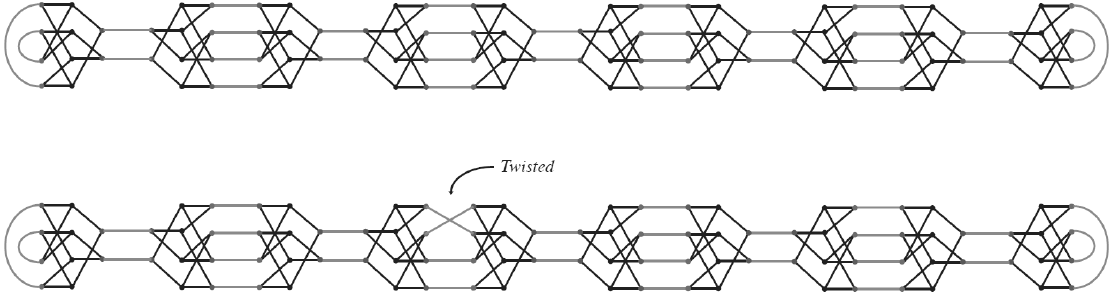}
 \caption{Miyazaki graph $M_4$ and
  twisted Miyazaki graph $MT_{4,2}$.}
 \label{Fig3}
\end{figure}

Similarly, we define a twisted Miyazaki graph $MT_{4,2}$ that is
identical with $M_4$ up to the 4th pair of bold knots: here two
connections are twisted (cf. Fig.~\ref{Fig3}). One can easily
generalize this construction to $MT_{n,k}$ with arbitrary
$n,k\in\mathbb N$ and $k\leq n$.

Distinguishing $M_n$ from $MT_{n,k}$ is a hard problem for graph
isomorphism software and is widely used as a benchmark test. The
following table shows the running times of our program on the
Miyazaki graphs of different size.

\vskip+0.2cm
\begin{center}
\extrarowheight=1pt
\begin{tabular}{| c || c | c | }
\hline
 \ Size \ & \ Matrix \ & \ Computation \ \\
 $n$& \ Dimensions \ & \ Time in sec \ \\ \hline
 10 & $200\times200$ & 9 \\ \hline
 15 & $300\times300$ & 26 \\ \hline
 20 & $400\times400$ & 60 \\ \hline
 25 & $500\times500$ & 133 \\ \hline \hline
 30 & $600\times600$ & 1069 \\ \hline
 35 & $700\times700$ & 1940 \\ \hline
 40 & $800\times800$ & 5755 \\
\hline
\end{tabular}
\end{center}
\vskip+0.2cm

\noindent \textbf{Remark.} The determinants of the original
connection matrices of Miyazaki graphs and their twisted
counterparts are zero, so there is no fast computation possible
using the inverse matrices. Therefore, we changed the
corresponding connection matrices inserting~3 as diagonal
elements. Due to this, for Miyazaki graphs of sizes $n\leq 25$
faster computation was available. This explains the sudden rise in
computation time for size 30 and more.

\subsection{Block designs: projective planes}

A projective plane of order $n$ is an incidence structure on
$n^2+n+1$ points, and equally many lines, (i.e., a triple
$(P,L,I)$, where $P$, $L$ and $I$ are disjoint sets of the points,
the lines and the incidence relation with $I\subset P\times L$ and
$|P| = |L| = n^2 + n + 1$), such that:
\begin{enumerate}[(i)]
 \item %
for all pairs of distinct points $p, p' \in P$, there is exactly
one line $ l\in L$ such that $(p, l)\in I$, $(p', l)\in I$;
 \item %
for all pairs of distinct lines $l, l'\in L$, there is exactly
one point $p\in P$ such that $(p, l)\in I$, $(p, l')\in I$;
 \item %
there are four points such that no line is incident with more
than two of these points.
\end{enumerate}

\begin{figure}[h]
 \centering
 \includegraphics[width=0.9\textwidth]{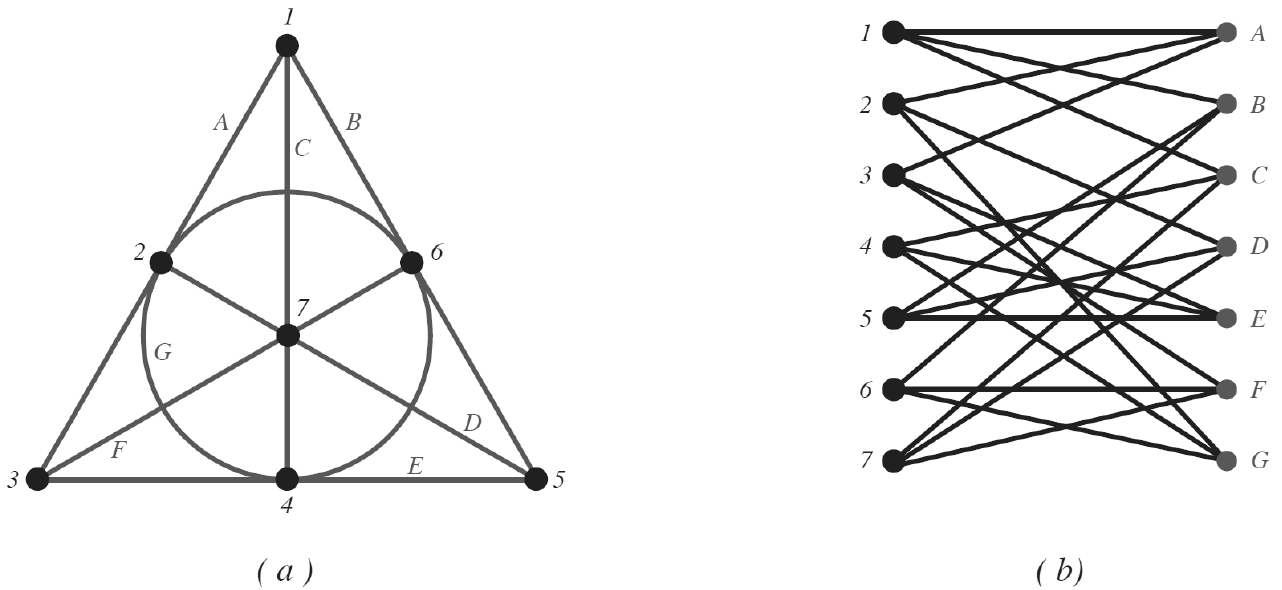}
 \caption{The least projective plane: the Fano plane.}
 \label{Fig4}
\end{figure}

The smallest projective plain is the Fano plane of order $2$ shown
in Fig.~\ref{Fig4}(a). It is also the only plane of this order.

We consider the corresponding incidence graphs of different
projective plains and try to solve the graph isomorphism problem
for them. The incidence graph of the Fano plane is shown in
Fig.~\ref{Fig4}(b). In general, distinguishing  projective planes
of the same order by their incidence graphs poses a difficult
challenge for graph isomorphism algorithms. The web page of
Moorhouse \cite{Moor} offers a collection of known projective
planes.

Since the computation in Wolfram's Mathematica 8 takes very long time for such large graphs, the algorithm was implemented in the C programming language. To 
calculate very large numbers while computing $\mathfrak A_G$ or $\mathfrak C_G$ the chinese remainder theorem was used. Each $d^G_{i,j,k}=|A_G^{i,j}|\mbox{\it \ mod }{p_k}$ 
was computed modulo a prime $43969\leq p_k\leq44699$ ($l\in\{1,2,\dots,72\}$) and the result was calculated by the formula
$$ |A_G^{i,j}|=\sum_{k=1}^{72}d^G_{i,j,k}m_ks_k, $$
where $m_k=m/p_k$, $m=p_1\cdots p_{72}$, and $s_k=m_k^{-1}\mbox{\it \ mod }{p_i}$.

So, instead of very large numbers, we have to manipulate with
lists of relatively small numbers: each $|A_G^{i,j}|$ corresponds
to a list of 72 natural numbers that we denote by
$$
 H_G^{i,j,k}=\big\{d^G_{i,j,k} \mid 1\leq k\leq72\big\}.
$$
In fact, it suffices to compute each $|A_G^{i,j}|\mbox{\it \ mod }{p_1}$
(modulo one sufficiently large prime). If the sorted corresponding
lists are differenent, $H_G^{i,j,1}\neq H_{G'}^{i,j,1}$, where
$$
 H_G^{i,j,k}=\mbox{\it \ Sort }
 \Big(\big\{|A_G^{i,j}|\mbox{\it \ mod }{p_k} \mid 1\leq k\leq72
 \big\}\Big),
$$
we can be sure that the graphs $G$ and $G'$ are not isomorphic, otherwise 
  we obtain two weighted graphs $(d^G_{i,j,k})_{i,j=1}^m$ and
$(d^{G'}_{i,j,k})_{i,j=1}^m$ with $m=2(n^2+n+1)$, where $n$ is the
order of the projective plane, and repeat the whole algorithm for
them. If $H_G^{i,j,k}=H_{G'}^{i,j,k}$ for each $1\leq k\leq72$ in
the second iteration, the graphs are \textit{conjectured} to be
isomorphic.

In general, if $H_G^{i,j,k}=H_{G'}^{i,j,k}$ for each $1\leq
k\leq72$, we can consider the weighted graphs as above and proceed
with the iteration, but we need a halting criteria to decide when
the two graphs are isomorphic. In our experiments, the differences
of most non-isomorphic graphs were discovered in the first
iteration step, however for some projective planes two iterations
were needed.

\vskip+0.2cm
\begin{center}
\extrarowheight=1pt
\begin{tabular}{| c || c | c | c | c |}
 \hline
 \ Order \ & No. of& \ Matrix \ & \ Iteration \ & \ Computation \ \\
 $n$&planes& \ dimensions \ & \ steps \ & \ time in sec \ \\ \hline
 9 & 4 & $182\times182$ & 2 & 63 $\cdot2$=126 \\ \hline
 11 & 1 & $266\times266$ & 2 & 162$\cdot$2=324 \\ \hline
 13 & 1 & $366\times366$ & 2 & 467$\cdot$2=934 \\ \hline
 16 & 22 & $546\times546$ & 1 & 5460 \\ \hline
 25 & 193 & $1302\times1302$ & 1 & 52080 \\ \hline
\end{tabular}
\end{center}
\vskip+0.2cm

Note that in some cases (such as for $n=11$, $13$, $17$, $19$, or
$23$) there is only one real projective plane, however it is
self-dual.

For $n=27$, the determinants for the $1415\times1415$ matrices
were calculated on a parallel machine with 20 Intel Xeon 2.80\,GHz
CPUs\footnote{The computations were carried out by Levan Kasradze.}. While
the computations for basic graphs took relatively reasonable time,
the computations for their duals lasted over twenty times as much
for one iteration step (see the table below).  

\vskip+0.2cm
\begin{center}
\extrarowheight=1pt
 \begin{tabular}{| c | c | c |}
 \hline
 Andre \ &Hering \ & \ Sherk \ \\ \hline \hline
 $15584.9+89377.6$ & \ $38125.7+9757.26$ \ & \ $44448.8+9726.34$ \\
 104962.5 & 47882.96 & 54175.14 \\ \hline \hline
 Andre dual \ &Hering dual \ & \ Sherk dual \ \\ \hline \hline
 944676 & \ 976157 \ & \ 964906 \\ \hline
\end{tabular}
\end{center}
\vskip+0.2cm

\vskip+0.2cm
\begin{center}
\extrarowheight=1pt
 \begin{tabular}{| c | c |}
 \hline
 Flag 4 \ & \ Flag 6 \ \\ \hline \hline
 $35047.3+9771.57$ \ & \ $34881+9773.86$ \\
 44818.87 & 44654.86 \\ \hline \hline
 Flag 4 Dual \ & \ Flag 6 Dual \ \\ \hline \hline
 984884 \ & \ 1036820 \\ \hline
\end{tabular}
\end{center}
\vskip+0.2cm

\section{Conclusions and Open Questions}

In this paper, we have developed an $O(n^5)$ polynomial-time
algorithm that computes graph invariant for graphs with $n$
vertices. The main approach is based on random walks on graphs
with the probability to stay in the actual node. Due to this, we
generate a set of $n(n-1)/2$ quadratic polynomials of one
variable. Comparing these sets we can distinguish different
graphs. In some cases, two iterations of the algorithm are needed.
The experimental results on some hard graphs (Miyazaki graphs as
special F\"urer gadgets and point-line incidence graphs of finite
projective planes of higher degrees) show that our system can
distinguish non-isomorphic graphs of this kind in reasonable time.
It is a matter of further research to prove (or disprove) that the
given method is a full graph invariant and to investigate the
question why there are relative break-ins in the computational
time for some point-line incidence graphs of {\it dual} finite
projective planes of high order.


\end{document}